\documentstyle[12pt]{article}
\begin{document}

\newcommand{\ra}{\rightarrow}
\newcommand{\ko}{K^0}
\newcommand{\be}{\begin{equation}}
\newcommand{\ee}{\end{equation}}
\newcommand{\bea}{\begin{eqnarray}}
\newcommand{\eea}{\end{eqnarray}}
\newcommand{\sqt}{\ft{1}{\sqrt{2}}\sin\phi\,}
\newcommand{\dtp}{\frac{\partial}{\partial\theta^+}}
\newcommand{\dtm}{\frac{\partial}{\partial\theta^-}}
\newcommand{\dtbp}{\frac{\partial}{\partial\bar\theta^+}}
\newcommand{\dtbm}{\frac{\partial}{\partial\bar\theta^-}}
\newcommand{\ft}[2]{{\textstyle\frac{#1}{#2}}}
\newcommand{\wt}{\widetilde}
\newcommand{\wh}{\widehat}
\def\vu{\varepsilon}
\def\vf{\varphi}
\def\a{\alpha}
\def\b{\beta}
\def\d{\delta}
\def\e{\epsilon}
\def\f{\phi}
\def\g{\gamma}
\def\h{\eta}
\def\k{\kappa}
\def\l{\lambda}
\def\m{\mu}
\def\n{\nu}
\def\o{\omega}
\def\p{\pi}
\def\q{\theta}
\def\r{\rho}
\def\s{\sigma}
\def\t{\tau}
\def\u{\upsilon}
\def\x{\xi}
\def\z{\zeta}
\def\G{\Gamma}

\newcommand{\del}{\partial}
\newcommand{\delb}{\bar{\partial}}

\renewcommand{\theequation}{\thesection.\arabic{equation}}
\newcommand{\eqn}[1]{eq.(\ref{#1})}
\renewcommand{\section}[1]{\addtocounter{section}{1}
\vspace{5mm} \par \noindent
  {\bf \thesection . #1}\setcounter{subsection}{0}
  \par
   \vspace{2mm} } 
\newcommand{\sectionsub}[1]{\addtocounter{section}{1}
\vspace{5mm} \par \noindent
  {\bf \thesection . #1}\setcounter{subsection}{0}\par}
\renewcommand{\subsection}[1]{\addtocounter{subsection}{1}
\vspace{2.5mm}\par\noindent {\em \thesubsection . #1}\par
 \vspace{0.5mm} }
\renewcommand{\thebibliography}[1]{ {\vspace{5mm}\par \noindent{\bf
References}\par \vspace{2mm}}
\list
 {\arabic{enumi}.}{\settowidth\labelwidth{[#1]}\leftmargin\labelwidth
 \advance\leftmargin\labelsep\addtolength{\topsep}{-4em}
 \usecounter{enumi}}
 \def\newblock{\hskip .11em plus .33em minus .07em}
 \sloppy\clubpenalty4000\widowpenalty4000
 \sfcode`\.=1000\relax \setlength{\itemsep}{-0.4em} }
{\hfill{VUB/TENA/96/07, hep-th/9610103}}
\vspace{1cm}
\begin{center}
{\bf THE GEOMETRY OF SUPERSYMMETRIC $\sigma$-MODELS\footnote{Contribution to
the proceedings of the workshop {\it Gauge Theories, Applied Supersymmetry and
Quantum Gravity, Imperial College, London, July 5-10, 1996}}}

\vspace{1.4cm}

ALEXANDER SEVRIN and JAN TROOST\footnote{Aspirant NFWO} \\
{\em Theoretische Natuurkunde, Vrije Universiteit Brussel} \\
{\em Pleinlaan 2, B-1050 Brussel, Belgium} \\
\end{center}
\centerline{ABSTRACT}

\vspace{- 4 mm}  

\begin{quote}\small
We review non-linear $\sigma$-models with (2,1) and (2,2) supersymmetry.
We focus on off-shell closure of the supersymmetry algebra and give a
complete list of $(2,2)$ superfields.
We provide evidence to support the conjecture that {\it all}
$N=(2,2)$ non-linear $\sigma$-models can be described
by these fields. This in its turn  leads to interesting consequences about the
geometry of
the target manifolds. One immediate
corollary of this conjecture is the existence of a potential for hyper-K\"ahler
manifolds, different
from the K\"ahler potential, which does not
only allow for the computation of the metric, but of the three fundamental
two-forms as well.
Several examples are provided:
WZW models on $SU(2)\times U(1)$ and $SU(2)\times SU(2)$ and
four-dimensional
special hyper-K\"ahler manifolds.
\end{quote}
\baselineskip17pt
\addtocounter{section}{1}
\section{Introduction and conclusions}
Non-linear $\sigma$-models with more than one supersymmetry are the
building blocks for stringtheories. For $N\geq(2,2)$ no complete off-shell
formulation of these models has been given. An off-shell realization is
desirable as it gives a manifest model independent description of the
supersymmetry, facilitates computations, makes the geometry (more)
obvious and finally it allows for the construction of the T-duals,
keeping the extended supersymmetry manifest.
In this paper, which is in part
a review \cite{st}, we investigate the (2,2) case. A complete
classification of (2,2) superfields exists: there are no other
superfields than chiral, twisted chiral \cite{GHR} and semi-chiral
\cite{BLR} ones. We provide several arguments to support the claim that
this is sufficient to describe {\it all} $(2,2)$ $\sigma$-models.
The central
object is the commutator of the left and right complex structures.
Its kernel is parametrized by chiral and twisted chiral coordinates and
correspond to ``K\"ahler-like'' directions. The complement
of the kernel is parametrized
using semi-chiral coordinates and can be
viewed as a deformation of a hyper-K\"ahler manifold.

If this conjecture turns out to be true, then one gets that the geometry of
a large class of complex manifolds is encoded in a potential, which allows
for the computation of the metric, torsion and complex structures. An
immediate corollary of this would be that for hyper-K\"ahler manifolds
there should exist a potential, not necessarily equivalent to the K\"ahler
potential which allows not only for the computation of the metric but of
the three complex structures as well!

Our results open several potentially interesting applications. A
systematic study of the T-duals along the lines of \cite{kiritsis} should
be done. Another point which deserves interest is the systematic
study of (2,2), (2,1) and (2,0) strings. Up to now, the only $N=2$ strings
studied are those described solely by chiral fields \cite{ov} and those
described by chiral and twisted chiral fields \cite{ch}. As will be shown in this
paper, very different choices can be made for the complex structures and it would
be interesting to know how the geometry of $(2,2)$ strings depends on this.
We presently investigate the geometry of $N=2$ strings with semi-chiral
fields.
Such a study could be relevant for the recent proposals in \cite{emil}
relating the $D=11$ membrane to type IIB stringtheory.
\setcounter{equation}{0}
\section{$N=(2,1)$ non-linear $\sigma$-models in superspace}
Omitting the dilaton term, a supersymmetric
non-linear $\sigma$-model in $N=(1,1)$ superspace is given by\footnote{
We take $D\equiv\frac{\del}{\del \q}+\q\del$ and
$\bar D \equiv \frac{\del}{\del \bar \q}+\bar\q\delb$,
with $\partial\equiv\frac{\partial}{\partial z}$ and
$\bar\partial\equiv\frac{\partial}{\partial{\bar z}}$.}
\begin{eqnarray}
{\cal S}=\int d^2 x d^2 \q\left(g_{ab}+b_{ab}\right)D\f^a\bar D
\f^b,\label{staction}
\end{eqnarray}
The metric on the target manifold is $g_{ab}$ and $b_{ab}=-b_{ba}$ is a
potential for the
torsion,
$T_{abc}\equiv-\frac 3 2 b_{[ab,c]}$.
A second, left-handed supersymmetry is of the form
\begin{eqnarray}
\d \f^a=\varepsilon J^a{}_bD\f^b\label{sus1}.
\end{eqnarray}
The action, eq. (\ref{staction}), is invariant provided\footnote{By
$\nabla^\pm$ we denote
covariant differentiation using the
$\G^a_{\pm bc}\equiv  \left\{ {}^{\, a}_{bc} \right\}
\pm T^a{}_{bc}$ connection.}
$ \nabla_c^+J^a{}_b=0$ and $J_{ab}=-J_{ba}$ hold.
One obtains the standard supersymmetry algebra if $J$ obeys
$J^2=-{\bf 1}$ and $N^a{}_{bc}[J]=0$,
with the Nijenhuis tensor given by
\be
N^a{}_{bc}[J]\equiv J^d{}_{[b}J^a{}_{c],d}+J^a{}_dJ^d{}_{[b,c]}.
\label{Nij2}
\ee
In other words $J$ is a  complex structure which is covariantly constant and
for which the metric is hermitean. This can easily be put in $(2,1)$
superspace. We choose a coordinate system such that the non-vanishing
components of
$J$ are
$J^\alpha {}_\beta=i\delta^\alpha _\beta$ and
$J^{\bar\alpha} {}_{\bar\beta}=-i\delta^\alpha _\beta$. Hermiticity is equivalent
to $g_{\alpha \beta}=g_{\bar\alpha \bar\beta}=0$.
The constancy of $J$ implies
\begin{eqnarray}
\Gamma^{\bar\alpha} _+{}_{\beta c}=0\Rightarrow T_{\alpha \beta\gamma}=0,\quad
(g_{\bar\gamma[\alpha }- b_{\bar\gamma[\alpha })_{,\beta]}=0, \label{conJ}
\end{eqnarray}
where we used that one can always gauge $b_{\alpha \beta}$ to zero.
Eq. (\ref{conJ}) implies that locally, metric and
torsion can be expressed in terms of a vector potential $k$:
\begin{eqnarray}
g_{\alpha \bar\beta}=\frac 1 2 (k_{\alpha ,\bar\beta}+k_{\bar\beta,\alpha
}),  \quad  b_{\alpha \bar\beta}=-\frac 1 4 k_{[\alpha
,\bar\beta]}\Rightarrow
T_{\alpha \beta\bar\gamma}=-\frac 1 4 (k_{\alpha ,\beta}-
k_{\beta,\alpha })_{,\bar\gamma} \label{defk}
\end{eqnarray}
The vectorfield $k_a$
is determined
modulo
$k_\alpha \simeq k_\alpha + f_\alpha +ig_{,\alpha }$,
where $f_{\alpha ,\bar\beta}=0$ and $g$ is a real function.
Introducing now a second Grassman coordinate $\hat\theta$ and denoting the
derivatives by ${\wh D}\equiv \frac{\partial\ }
{\partial{\hat\theta}}-\wh\theta\partial$, we get the action in $(2,1)$
superspace:
\begin{eqnarray}
{\cal S}=\frac 1 2 \int d^2z d^2\theta d\hat\theta\left(
k_\alpha \bar D \phi^\alpha -k_{\bar\alpha }\bar D \phi^{\bar\alpha }
\right),
\end{eqnarray}
where $\phi$ are $(2,1)$ chiral fields:
$\widehat{D}\phi^\alpha =-D\phi^\alpha$,
$\widehat{D}\phi^{\bar\alpha} =D\phi^{\bar\alpha}$.
\setcounter{equation}{0}
\section{$N=(2,2)$ non-linear $\sigma$-models in superspace}
\subsection{(2,2) supersymmetry}
We turn back to the action eq. (\ref{staction}) and consider a second,
non-chiral supersymmetry:
$\d \f^a=\varepsilon J^a{}_bD\f^b+\bar\varepsilon
\bar J^a{}_b\bar D\f^b$.
Requiring invariance and a standard {\it on-shell} $N=(2,2)$ supersymmetry
algebra gives that both $J$ and $\bar J$ are covariantly constant
($J$ w.r.t. the
$\Gamma_+$ and $\bar J$ w.r.t. the
$\Gamma_-$ connection) complex structures such that the
metric is hermitean for both. The only off-shell non-closure comes from the
commutator of the right-handed with the left-handed supersymmetry:
\begin{eqnarray}
[\delta(\varepsilon ),\delta (\bar\varepsilon)]\phi^a =
\varepsilon\bar\varepsilon[J,\bar J]^a{}_b(
D\bar D \f^b + \G^b_{-cd}  D\f^d\bar D \phi^c).
\end{eqnarray}
One recognizes the equation of motion for $\phi$ preceeded by the commutator of
$J$ and $\bar J$. This leads to the important observation
that the algebra closes
off-shell in the direction of $\ker [J,\bar J]$, which hints towards the
possibility that $\ker [J,\bar J]$ can be described without the introduction of
additional auxiliary fields while the complement of $\ker [J,\bar J]$ will
need  auxiliary fields.
\subsection{N=(2,2) superfields}
In addition to the $(1,1)$ superspace coordinates, we introduce
two new fermionic coordinates\footnote{In the
literature, one often finds the
coordinates $\q^+$ and $\q^-$. They are related to our coordinates by
$\q^\pm=\frac{1}{\sqrt{2}}(\q\pm\wh\q)$},
$\wh\theta$ and $\wh{\bar\theta}$. The derivatives are given by
\begin{eqnarray}
\wh D\equiv\frac{\del}{\del \wh \q}-\wh
\q\del,\qquad \wh {\bar D}\equiv\frac{\del}{\del \wh{\bar \q}}-\wh
{\bar \q}\delb.
\end{eqnarray}
The lagrange density in $(2,2)$ superspace can only be
a function of scalar fields, so the dynamics will be largely determined by the
choice of superfields. Constraints \cite{st}
on a set of general superfields $\phi^a$,
$a\in\{1,\cdots , n\}$, are of the form $\wh D\phi^a=iJ(\phi)^a{}_bD\phi^b$.
Integrability ($\wh D^2 =-\partial$) of this requires $J$ to be a complex
structure. Imposing additional constraints of opposite chirality
$\wh{ \bar D }\phi^a=i \bar{J}(\phi)^a{}_b \bar D\phi^b$ require not only that
$\bar J$ is a complex structures but from  $\{ \wh D, \wh{\bar D}\}=0$,
impose that $J$ and $\bar J$ commute as
well.
Constraining both chiralities reduces the degrees of freedom of a general
superfield to those of an $N=(1,1)$ field. One shows \cite{st} that through an
appropriate coordinate transformation, $J$ and $\bar J$ can be diagonalized
simultanously. The eigenvalues, $\pm i$, can be combined in four different
ways, yielding the basic superfields:
\begin{enumerate}
\item chiral field $\Phi$ and anti-chiral field $\bar \Phi$:
\begin{eqnarray}
\wh{D}\Phi\, =\, -D\Phi,\quad \wh{\bar D}\Phi\,=\, -\bar D\Phi,\qquad
\wh{D}\bar\Phi\, =\, +D\bar\Phi,\quad \wh{\bar D}\bar\Phi\,=\, +
\bar D\bar\Phi.
\end{eqnarray}
\item twisted chiral field $\Phi$ and twisted anti-chiral field $\bar \Phi$
\cite{GHR}:
\begin{eqnarray}
\wh{D}\Phi\, =\, -D\Phi,\quad \wh{\bar D}\Phi\,=\, +\bar D\Phi,\qquad
\wh{D}\Phi\, =\, +D\Phi,\quad \wh{\bar D}\Phi\,=\, -\bar
D\Phi.\label{basdef}
\end{eqnarray}
\end{enumerate}
There is only one other type of superfield, the {\it semi-chiral
superfield} \cite{BLR}, in which only one chirality is constrained. An analysis of the
integrability conditions and the requirement that in
the end we want a $\sigma$-model forces us to take them in pairs,
such that the constraints on each member are of opposite chirality.
So contrary to the previous fields which correspond to two real dimesions,
a semi-chiral multiplet describes four real dimensions.
Each member of the pair contains now {\it two} $N=(1,1)$ superfields, one of
which will be auxiliary. We come back to a detailed study of this in the following
section.
\subsection{$N=(2,2)$ non-linear $\sigma$-models in superspace}
An obvious question is whether all non-linear $\sigma$-models can be
described using the fields mentioned above. When $[J,\bar J]=0$, the
model can be described using chiral and twisted chiral fields \cite{GHR},
which parametrize $\ker (J-\bar J)$ and $\ker (J+\bar J)$ resp., where
$\ker[J,\bar J]=\ker (J-\bar J)\oplus \ker (J+\bar J)$. Such manifolds
have a product structure: $\Pi\equiv J\bar J$ with $\Pi^2={\bf 1}$. The
projection operators $P_\pm\equiv\frac 1 2 ({\bf 1}\pm \Pi)$ project on
$\ker (J\pm\bar J)$, where each of the subspaces is K\"ahler. Introducing a real
potential $K$, function of these fields and denoting
the chiral and twisted chiral directions by indices $\alpha$ and $\mu$
resp., one easily computes the vector
$k$ introduced in eq. (\ref{defk}): $k_\alpha =-K_\alpha$ and $k_\mu= K_\mu$.
If we write the potential with subindices, we mean the derivatives of the
potential w.r.t. those fields.

Remains the case where $[J,\bar J]\neq 0$.
An important result \cite{martinnew} states
that $\ker (J-\bar J)$ and $\ker (J+\bar J)$ are always integrable to chiral
and twisted chiral fields resp. This leaves us with the subspace where $\ker[J,\bar J]$
is non-degenerate, which we expect to be parametrized by semi-chiral
fields. As one semi-chiral multiplet corresponds to four real dimensions,
the complement of $\ker[J,\bar J]$ needs to have a dimension which is a
multiple of four. One can show \cite{st} that this is indeed the case. Let
us restrict our attention to manifolds where $\ker[J,\bar J]=\emptyset$. If we
diagonalize one of the complex structures, then we get the following structure:
\begin{eqnarray}
J=i\left(\begin{array}{cc}{\bf 1}&0\\0&-{\bf 1}
\end{array}\right),\quad \bar J=\left(\begin{array}{cc}a&b\\-b^{-1}(1+a^2)&-b^{-1}ab
\end{array}\right),\label{cs2o}
\end{eqnarray}
and $a^2\neq -{\bf 1}$. Both $a$ and $b$ have to satisfy several requirements
\cite{st}. It remains to be shown that the general solution to these equations
is indeed provided by a semi-chiral parametrization. However, the previous and
following arguments, together with several explicitely worked out
examples, support this claim. Before turning to the semi-chiral
parametrization, we point out an interesting feature:
$a=0$ corresponds to hyper-K\"ahler
manifolds. In this way one can view a generic manifold with
$\ker[J,\bar J]=\emptyset$ as a deformation of a hyper-K\"ahler manifold.

We take $n$ semi-chiral multiplets $\{ \phi^\alpha , \phi^{\bar\alpha },
\eta^{\wt\alpha }, \eta^{\wt{\bar\alpha }}\}$.
Through a coordinate transformation, one can always reduce the defining relations
of a semi-chiral multiplet \cite{st} to:
\begin{eqnarray}
\widehat{D}\f^\a=-D\f^\a,\quad
\widehat{D}\f^{\bar \a}=D\f^{\bar \a},\quad
\widehat{\bar D}\eta^{\wt{ \a}}=\bar D \eta^{\wt{
\a}},\quad
\widehat{\bar D}\eta^{\wt{\bar
\a}}=-\bar D \eta^{\wt{\bar \a}}.\label{defsc}
\end{eqnarray}
Taking an arbitrary real potential $K(\phi,\eta,\bar\phi,\bar\eta)$,
which is determined modulo
a generalized K\"ahler transformation
$K\simeq K+f(\phi ) +
g(\eta) + \bar f(\bar\phi) + \bar g (\bar\eta)$,
we pass to $(1,1)$ superspace:
\begin{eqnarray}
&&{\cal S} =\int\,d^2z d^2\theta d^2{\hat{\theta}}\, K=
\int\,d^2zd^2\theta\Big\{ \chi^T L\psi-\bar
D\h^TPLPD\f-\nonumber\\
&&\chi^T(L\psi-PL\bar  D\f-2P\wt M \bar D \h)-
(\chi^T L+D\h^TLP-2D\f^T P M )\psi\Big\},
\end{eqnarray}
where $L$, $M$ and $\wt M$ are $2n\times 2n$ matrices
\bea
L\equiv \left( \begin{array}{cc} K_{\wt\a\b} & K_{\wt\a\bar\b}\\
K_{\wt{\bar\a}\b} & K_{\wt{\bar\a}\bar\b}\end{array}\right), \
\wt M\equiv \left( \begin{array}{cc} 0 & K_{\wt\a\wt{\bar\b}}\\
K_{\wt{\bar\a}\wt\b} & 0\end{array}\right),\
M\equiv \left( \begin{array}{cc} 0 & K_{\a{\bar\b}}\\
K_{{\bar\a}\b} & 0\end{array}\right),\label{defmat}
\eea
and $\f$ and $\h$ are $2n\times 1$ matrices while $P$ is a constant $2n\times 2n$ matrix:
\bea
\f\equiv\left( \begin{array}{c} \f^\a \\ \f^{\bar \a}\end{array}\right),
\qquad
\h\equiv\left( \begin{array}{c} \eta^{\wt \a} \\ \eta^{\wt{\bar
\a}}\end{array}\right),\qquad P\equiv \left( \begin{array}{cc}{\bf 1} & 0\\
0 & -{\bf 1}\end{array}\right) .
\eea
Assuming that $L$ is invertible, one can eliminate the auxiliary fields $\chi\equiv
\wh D \eta$ and $\psi\equiv\wh{\bar D}\phi$
through their e.o.m. and one gets the
second order action,
\begin{eqnarray}
{\cal S} &=& \int d^2zd^2\theta\Big\{-D\phi^T
PL^TP\bar D\eta +\nonumber\\
&&(D\h^TL+2D\f^T  M ) P L^{-1} P ( L\bar D \f+ 2  \wt M \bar D \h )\Big\},
\label{ac1o}
\end{eqnarray}
from which both the metric and the torsion potential can be read off.
$J$ can be diagonalized through the coordinate
transformation $\phi^a\rightarrow \varphi^a=\phi^a$, $\eta^a\rightarrow
\varphi_a=K_a$. Using this, we get $k$ defined
in eq. (\ref{defk}). In the original semi-chiral coordinates we have
${\wt k} =LPL^{-1}P{\wt K}$ and $k=2ML^{-1}{\wt k}$,
where
\begin{eqnarray}
{\wt K}\equiv\left(\begin{array}{c}K_{\wt\alpha}\\K_{\wt{\bar\alpha}}
\end{array}\right),\quad
{\wt k}\equiv\left(\begin{array}{c}k_{\wt\alpha}\\k_{\wt{\bar\alpha}}
\end{array}\right),\quad
{k}\equiv\left(\begin{array}{c}k_{\alpha}\\k_{{\bar\alpha}}
\end{array}\right).
\end{eqnarray}
Parmetrizing rows as $(\phi,\eta)$, one also gets the complex structures:
\bea
J=\left(\begin{array}{cc}iP&0\\
2iL^{-1}{}^T P M  & i L^{-1}{}^T
P  L^T
\end{array}\right),\
\bar J =\left(\begin{array}{cc}-iL^{-1}PL& 2iL^{-1} \wt M P\\
0&-iP \end{array}\right).\label{cs1}
\eea
Requiring the metric to be non-degenerate implies that $\ker [J,\bar
J]=\emptyset$!

The necessary and sufficient conditions to have a semi-chiral description of
a hyper-K\"ahler manifold are
\begin{eqnarray}
&&L^{-1}PLP+PL^{-1}PL=4L^{-1}P\wt ML^{-1T}MP\nonumber\\
&&\{ P, L^{-1T}MPL^{-1}\}=\{P,L^{-1}P\wt M L^{-1T}\}=0.
\end{eqnarray}
Restricting ourselves to $d=4$, we find that the two latter eqs. are
trivially satisfied while the former becomes:
$ |K_{\phi\eta}|^2+|K_{\phi\bar \eta}|^2=2K_{\eta\bar \eta}K_{\phi\bar\phi}$.
It is
known that a 4-dimensional K\"ahler manifold is hyper-K\"ahler iff. the
K\"ahler potential satisfies the Monge-Amp\`ere equation. So a concrete
way to test our hypothesis would be to show that somehow the previous equation
is equivalent to the Monge-Amp\`ere
equation. Looking at arbitrary dimensions, we see that we get a full set of
equations similar to those obtained in \cite{ulf}. The problem in proving
our conjecture is essentially that while it is very easy to pass from
semi-chiral coordinates to coordinates where one of the complex structures
is diagonal, the reverse is not true. We are presently studying this
particular point.
\setcounter{equation}{0}
\section{Examples}
\subsection{Wess-Zumino-Witten models}
WZW-models on even dimensional groups are particular examples of (2,2)
$\sigma$-models \cite{usold}. The complex structures are easily
characterized by their action on the Lie algebra: they are almost
completely determined by a Cartan decomposition. The complex structure has
eigenvalue $+i$ and $-i$ on generators corresponding with positive
and negative roots resp. The only freedom left is the action of the complex
structure on the Cartan subalgebra (CSA). Except for the requirement that the
structure maps the CSA bijectively to itself, no further conditions have to be
imposed. One has that, except for $SU(2)\times U(1)$,
$[J,\bar J]\neq 0$ \cite{RSS}. Choosing for $SU(2)\times U(1)$
the left and right complex structures
so that they differ by a sign on the CSA, the
complex structures commute and the model can be parametrized by a chiral
$\phi$, and a twisted chiral field $\chi$. The potential is given by
\cite{RSS}
\begin{eqnarray}
K=-\int^{\frac{|\chi |^2}{|\phi|^2}}\frac{d\zeta }{\zeta }\ln (1+\zeta )
\, +\, \frac 1 2 \left( \ln (\phi\bar\phi)\right)^2 .
\end{eqnarray}
If on the other hand we choose left and right complex structures to be
equal on the Lie algebra, then $\ker[J,\bar J]=\emptyset$ and we can
describe the model with one semi-chiral multiplet with potential \cite{st,martinnew}
\begin{eqnarray}
K=-\f \bar{\f} + \bar{\f} \bar{\eta} + \f \eta
-2 i \int^{\bar{\eta}-\eta} dx \, \ln ( 1+ \exp \frac{i}{2} x).
\end{eqnarray}
Finally, an interesting example where different multiplets occur, is
$SU(2)\times SU(2)$  \cite{st}. Choosing both complex structures to be equal on the
Lie algebra we get that $\ker[J,\bar J]$ is two-dimensional. The manifold
can be parametrized by one chiral field $\zeta$ and a semi-chiral multiplet.
The potential is explicitely given by:
\bea
K &=& -\zeta \bar{\zeta} + \zeta \bar{\phi} + \bar{\zeta} \phi + i \eta \zeta
   - i \bar{\eta} \bar{\zeta}  + i \bar{\eta} \bar{\phi} - i \eta \phi
   \nonumber \\
 &&  - i \int^{\bar{\phi}-\phi} dy \, \ln (1-\exp i y)
   - i \int^{\bar{\eta}-\eta} dy \, \ln (1-\exp i y) .
\eea
\subsection{Special hyper-K\"ahler manifolds}
As already mentioned, hyper-K\"ahler manifolds are an interesting class of
manifolds to test our conjecture. We present here the particular example of
four-dimensional special hyper-K\"ahler manifolds. They arise as the
scalar subsector of hypermultiplets in rigid $N=2$, $d=4$. The full
structure of these manifolds is explained elsewhere \cite{stef} and we
provide here just what is needed. The manifolds are parametrized by
coordinates $x$ and $v$ and the K\"ahler potential is given by $K_K=2i(
F_x\bar x-\bar F_{\bar x} x)+i(\bar v-v)^2(F_{xx}-\bar F_{\bar x\bar
x})^{-1}$, where $F(x)$ is a holomorphic prepotential. The first
fundamental 2-form is just the standard K\"ahler two form while the two
other ones are simply $\omega_2=2(dx\wedge dv+d\bar x\wedge d\bar v)$ and
$\omega_3=2i(dx\wedge dv-d\bar x\wedge d\bar v)$. The semi-chiral
parametrization is obtained through the coordinate transformation:
\begin{eqnarray}
x\rightarrow \phi=x,\quad v\rightarrow \eta=-2i(F_x+\bar F_{\bar
x})+x+\bar x+\frac{1-2i\bar F_{\bar x\bar x}}{N}v -
\frac{1-2iF_{xx}}{N}\bar v,
\end{eqnarray}
with $N\equiv i(F_{xx}-\bar F_{\bar x\bar x})$. The semi-chiral potential is
\begin{eqnarray}
K_{SC}&=&\frac 1 2 \eta\bar \eta+(\phi+\bar\phi)^2+4(F_\phi+\bar F_{\bar\phi})^2-
(2i(F_\phi+\bar F_{\bar\phi})+\phi+\bar\phi)\eta+\nonumber\\
&&(2i(F_\phi+\bar F_{\bar\phi})-\phi-\bar\phi)\bar\eta.
\end{eqnarray}
Using the formulas given
in section (3.3), one computes from the potential not only the metric but
the three complex structures as well.

\vspace{5mm}

\noindent{\bf Acknowledgments}
We thank  J. De Jaegher, B. de Wit, M. Ro\v{c}ek and S. Vandoren
for interesting discussions. This work
was supported in part by the European Commission TMR programme
ERBFMRX-CT96-0045 in which both authors are associated to K. U. Leuven.

\end{document}